\documentclass[pmlr,twocolumn,10pt]{jmlr}

\usepackage{booktabs}
\usepackage{threeparttable}

\jmlrproceedings{}{}
\jmlrworkshop{}
\jmlryear{2026}
\makeatletter
\renewcommand\ps@jmlrtps{%
  \let\@mkboth\@gobbletwo
  \def\@oddhead{}\let\@evenhead\@oddhead
  \def\@oddfoot{\@titlefoot}\let\@evenfoot\@oddfoot
}
\makeatother

\newcommand{\X}{\mathbf{X}}
\newcommand{\x}{\mathbf{x}}

\title[When Are Time-to-Event Models a Waste of Time?]{When are time-to-event
models a waste of time? \\
Bridging mixture cure models and positive-unlabeled learning for binary classification under right-censoring}

\author{%
 \Name{Qin Weng} \Email{wengqin.weng@duke.edu}\\
 \Name{Matthew M. Engelhard} \Email{matthew.engelhard@duke.edu}\\
 \addr Department of Biostatistics and Bioinformatics, Duke University,
 North Carolina, USA
}

\begin{document}

\maketitle

\begin{abstract}
In clinical settings, predicting binary outcomes is often complicated by right-censoring, which prevents us from distinguishing individuals in whom the event never occurs (i.e., \textit{negative}, or \textit{non-susceptible}) from those where it occurs after the censoring time (i.e., \textit{positive}, or \textit{susceptible}). To discriminate between these subpopulations, we can use  mixture cure models (MCMs), which model both the outcome probability over an infinite horizon and its time-to-event (TTE) distribution when observed. However, the TTE approach requires event timestamps, which can be unreliable, costly to obtain, or biased due to disparities in clinical care.  Alternatively, the task can be formulated as positive-unlabeled (PU) learning, in which individuals without an observed event are grouped as ``unlabeled'' rather than negative, recognizing that this group includes positives whose event occurred after censoring. Here we analyze the relationship between these frameworks and provide evidence-based recommendations for choosing between them. We begin by showing that both families optimize a shared likelihood but differ in how they constrain the probability of event occurrence prior to censoring among susceptible patients, which in PU learning is known as the \textit{labeling propensity}; and that an MCM may be viewed as a PU model whose labeling propensity is parameterized as a TTE distribution, which makes the model identifiable. In systematic simulations and two real-world clinical cohorts, we examine whether and when including the TTE component benefits binary classification. We show that MCMs are preferred in only one specific regime: under selective censoring where the exact features governing censoring cannot be identified. Otherwise, PU learning achieves equivalent performance while avoiding challenges and potential biases associated with the TTE approach.
\end{abstract}


\section{Introduction}
\label{sec:intro}

Building binary classification models in clinical contexts often involves predicting a long-term outcome, such as cancer recurrence or neurodevelopmental conditions. The goal is to 
discriminate between \textit{positives}, in whom the outcome (e.g., autism diagnosis, cancer recurrence) is eventually observed, and \textit{negatives}, in whom it is not.
However, in clinical settings with limited follow-up, 
we cannot distinguish between negatives and positives for whom the outcome takes place after censoring. 
For example, 
autism can be diagnosed well into adulthood,
therefore we are rarely certain of negative autism status in real-world studies with limited follow-up.

Mixture cure models (MCMs) are a natural, commonly used approach to model 
right-censored binary outcomes. An MCM posits two components: a binary classification model specifying whether the outcome will eventually occur, and a time-to-event (TTE) model specifying its TTE distribution when observed.
The former 
can be used directly to make predictions about binary status, while the latter translates timing information into evidence that helps distinguish positive (i.e., \textit{susceptible}) individuals, in whom the outcome occurs, though possibly after censoring; from negative (i.e., \textit{non-susceptible}) individuals, in whom it does not.
However, diagnosis times are frequently unavailable or unreliable in practice, as recorded timestamps can be confounded by administrative reporting delays \citep{schulz2021temporal}. Even when recorded times are trustworthy, the TTE model in MCMs remains vulnerable to model misspecification or poor fitting, thereby risking degrading the classification model rather than informing it. This raises a practical question: when does the TTE component help us predict binary outcomes effectively, and when is it unnecessary or even a liability?

Alternatively, the task can be formulated as positive-unlabeled (PU) learning.
Right-censored data has an inherent positive-unlabeled structure, since only
those patients with observed outcomes
within the observation window carry a confirmed positive label, while
all others remain unlabeled, since their outcome status over an infinite horizon is unknown \citep{toyabe2020positive}. PU learning methods can then learn
a binary classifier predicting positive/negative (i.e., susceptibility) status from patient features and observed diagnosis indicators
without explicitly incorporating a TTE model.
In exchange,
PU methods typically place additional assumptions on the underlying labeling mechanism that determines the probability of outcome occurrence prior to censoring among positive cases,
providing a structural constraint to help disentangle
positives and negatives
within the unlabeled set \citep{bekker2020learning}.

Although
both MCMs and PU learning have been extensively studied, to our knowledge no work has recognized the close connection between the two, placed them within a common analytic framework, or empirically investigated their relative merits for binary classification under right-censoring.
The importance of TTE model in this specific task has therefore never been isolated, leaving it unclear whether and how it facilitates effective binary classification. If incorporating timing information does not prove beneficial, simpler and more general PU learning methods would be preferred for their lower data requirements and broader applicability.

In this study, we aim to answer whether and when a TTE model should be incorporated to handle right-censoring in clinical outcome prediction. The paper is structured as follows. In Section~\ref{sec: method}, we present the mathematical setup and demonstrate how variants of MCMs and PU learning can be formulated within a unified framework and related to each other; a broader review of the two literatures is given in Appendix~\ref{app: review}. In Section~\ref{sec: simulation}, we present a simulation study that systematically compares these methods under varying conditions, including sample size, susceptibility rate, censoring rate, and the underlying feature-dependency for censoring and susceptibility. In Section~\ref{sec: real}, we further validate our findings in two real-world clinical datasets with both synthetic censoring and natural censoring. Finally, in Section~\ref{sec: con} we summarize whether and when the TTE component is beneficial, and condense the answer into a decision flowchart in Appendix~\ref{app: flow} that recommends a method given properties of the data at hand.

\section{Unified Framework for Censored Data Classification}\label{sec: method}

We formulate a unified framework to analyze the task of binary classification with right-censored data. We cast both MCMs and PU learning as different strategies for optimizing a shared objective, and show that MCMs can be viewed as a special variant of PU learning, with TTE modeling as a specific type of learning constraint.

\subsection{Setup}\label{sec: setup}

We consider a classification task that predicts a latent binary label, $Z \in \{0, 1\}$, from a $p$-dimensional feature vector $\X\in \mathbb{R}^p$. The main difficulty is that a proportion of positive cases are not labeled as such, because they are right-censored before the outcome is observed. In the training data, we do not directly observe $Z$; instead, we observe a binary variable $S \in \{0, 1\}$ indicating whether the outcome was observed; along with an observed time $Y\in \mathbb{R}^+$. The observed time is defined as $Y = \min(T, C)$, the earlier of the diagnosis time $T\in \mathbb{R}^+$ and the censoring time $C\in \mathbb{R}^+$. The indicator is set to $S=1$ only when an individual is positive $(Z=1)$ and the diagnosis occurs before censoring $(T\leq C)$; in all other cases, including negative individuals $(Z=0)$ and positive individuals censored before diagnosis $(T> C)$, we observe $S=0$.

Our goal is to estimate the outcome probability $p(\X)=\Pr(Z=1\mid\X)$ without directly accessing $Z$. A common approach is PU learning methods, which train a classifier using a basic dataset $\mathcal{D}_{binary} = \{(\x_i, s_i)\}_{i=1}^N$ that includes only features and diagnosis indicators.
In our specific setting, where the missing labels arise from right-censoring, additional TTE information may optionally be obtained. This enables us to work with a richer dataset $\mathcal{D}_{tte} = \{(\x_i, s_i, y_i)\}_{i=1}^N$, and to develop MCMs that jointly model both the outcome probability and the labeling process driven by censoring.

The right-censoring belongs to what \citet{bekker2020learning} term the single-training-set scenario of PU learning: the cohort is a single sample drawn i.i.d.\ from the population, and a subset of the positive individuals subsequently becomes labeled through diagnosis. A central concept in this framework is the labeling propensity, defined as
$$e(\X)=\Pr(S=1\mid Z=1,\X).$$
For simplicity, we may refer to the labeling propensity as the \textit{propensity} in subsequent discussions. We may then write the complete-data log-likelihood of observing indicators $S$ given their corresponding latent outcome labels $Z$:

\begin{equation}\label{eq: complete_lld}
    \begin{aligned}
l(p,e;z,&\mathcal{D}_{binary}) = \sum_{i=1}^n \Big[ s_iz_i\ln[p(\x_i)e(\x_i)] \\
  &+ (1-s_i)z_i\ln[p(\x_i)(1-e(\x_i))] \\
  &+ (1-s_i)(1-z_i)\ln[1-p(\x_i)] \Big].
    \end{aligned}
\end{equation}

While directly optimizing this log-likelihood is infeasible since $Z$ is latent, working from the indicator $S$ always leads to two explanations of being negative or right-censored, and nothing in the data $\mathcal{D}_{binary}$ can directly distinguish them. Therefore, every method has to impose some restriction on the propensity outside the complete-data likelihood in Equation~\ref{eq: complete_lld}. We divide strategies for fitting this log-likelihood into three groups according to the type of restriction: the propensity may be fixed to a constant across patients (Section~\ref{sec: scar}), or be only dependent with prior-known feature subspace (Section~\ref{sec: sar}), or can be explicitly modeled regarding the labeling process (Section~\ref{sec: mcm}). In the censored-data classification task, the first two groups require only $\mathcal{D}_{binary}$ and the third requires $\mathcal{D}_{tte}$.

\subsection{Constraint from Constant Propensity}\label{sec: scar}

The most widely used and most restrictive constraint is the ``selected completely at random'' (SCAR) assumption, under which all positive individuals share the same propensity $e(\X) = c$. A constant propensity makes $\Pr(S=1)$ proportional
to the latent outcome probability $\Pr(Z=1)$, which forms the basis for SCAR-based methods to reconstruct the population distribution from partially labeled data and converts the problem into a standard supervised learning task.

\cite{elkan2008learning} estimate $c$ as the average predicted
$\Pr(S=1\mid\X)$ over the labeled positives, which is valid
because SCAR renders them representative of all positives.
Writing $\Pr(S=1\mid\X)=c\cdot p(\x)$ gives the marginal
log-likelihood over $Z$ with $c$ as a fixed scaling parameter:

\begin{equation}\label{eq: marginal_lld_s}
\begin{aligned}
l(p;\mathcal{D}_{binary},c) = \sum_{i=1}^n
  & s_i\ln[c\cdot p(\x_i)] \\
  &+ (1-s_i)\ln[1-c\cdot p(\x_i)].
\end{aligned}
\end{equation}

An alternative strategy imposes the same SCAR restriction during training rather than post-hoc, by reweighting labeled and unlabeled instances so that the empirical loss remains unbiased \citep{du2014analysis}. This unbiased PU (uPU) objective is usually paired with robust losses \citep{du2015convex, kiryo2017positive}. When instantiated with the log-likelihood loss, this objective aligns seamlessly with our framework and recovers a reweighted form of the complete-data log-likelihood (shown in Appendix~\ref{app: scar}).

\subsection{Constraint from Feature Space}\label{sec: sar}

Relaxing the SCAR assumption enables more realistic modeling by allowing the labeling propensity to depend on patient features. However, replacing the constant propensity $c$ in the marginal log-likelihood (Equation~\ref{eq: marginal_lld_s}) with a feature-dependent propensity model $e(\x_i)$ results in an updated marginal log-likelihood with compromised model identifiability:

\begin{equation}\label{eq: marginal_lld_e}
    \begin{aligned}
l(p,e;\mathcal{D}_{binary}) = \sum_{i=1}^n
  & s_i\ln[e(\x_i)p(\x_i)] \\
  &+ (1-s_i)\ln[1-e(\x_i)p(\x_i)],
    \end{aligned}
\end{equation}

Here the models of outcome probability $p(\X)$ and propensity $e(\X)$ always appear together as a product. As a result, scaling one by a constant $c>0$ while scaling the other by $1/c$ -- or even swapping the two models entirely -- leaves the log-likelihood unchanged. This fundamental ambiguity makes the model unidentifiable: an unlabeled individual can always be equally well explained by assigning them a low outcome probability or a low propensity, respectively.

Therefore, additional assumptions and constraints become necessary. For instance, \citet{bekker2019beyond} relax SCAR to the Selected At Random (SAR) assumption: propensity is independent of outcome probability conditioned on the feature values. Based on that, they introduce the SARPU framework \citep{bekker2019beyond} as a generalized solution by restricting the feature set used for the propensity model to be a strict subset of the features used for the outcome model, thereby resolving the model's inherent unidentifiability issue.

\subsection{Constraint from TTE}\label{sec: mcm}

In our specific problem setting, the selective labeling process arises from the interplay between feature-dependent diagnosis times and censoring times. Therefore, MCMs provide a natural way to explicitly model the propensity from TTE predictions. Let $f(t)$ and $g(c)$ denote the densities of the latent diagnosis time $T$ and censoring time $C$, with corresponding survival functions $F(y\mid\x)=1-\int^y_0 f(t\mid\x)dt$ and $G(y\mid\x)=1-\int^y_0 g(c\mid\x)dc$, respectively. The propensity $e(\X)$ can then be expressed as the probability that a diagnosis occurs before censoring,
\begin{equation*}
    \begin{aligned}
e(\X) &= \Pr(T<C\mid Z=1,\X) \\
      &= 1 - \int F(y\mid\X)g(y\mid\X)\,dy,
    \end{aligned}
\end{equation*}
and the marginal log-likelihood given the extended TTE dataset is:
\begin{equation}\label{eq: marginal_lld_mcm}
    \begin{aligned}
l(p,f,g;&\mathcal{D}_{tte}) = \sum_{i=1}^n s_i\ln[p(\x_i)f(y_i\mid\x_i)G(y_i\mid\x_i)] \\
  &+ (1-s_i)\ln\big[p(\x_i)F(y_i\mid\x_i)g(y_i\mid\x_i) \\
  &\qquad\qquad\quad + (1-p(\x_i))g(y_i\mid\x_i)\big].
    \end{aligned}
\end{equation}
For simplicity, the censoring time is commonly assumed to be independent of the diagnosis times given covariates, therefore, the censoring term does not influence parameter estimation and can be safely omitted from the log-likelihood optimization.

\subsection{Relations Among Methods}\label{sec: relation}

We want to emphasize the high similarity between these method families. Particularly, marginalizing $Y$ out of Equation~\ref{eq: marginal_lld_mcm} returns exactly Equation~\ref{eq: marginal_lld_e}, with propensity model $e$ confined to the family induced by diagnosis time model $f$ and censoring time model $g$. Therefore, an MCM is essentially a SAR-PU model whose propensity is parameterized through the diagnosis and censoring time distributions rather than fit freely, so the TTE component acts as a constraint on propensity model $e$ rather than as an independent source of information about classifier $p$. The three groups therefore optimize the same objective under successively weaker restrictions on $e$: SCAR fixes its value, SARPU restricts the features it may depend on, and an MCM restricts its shape through TTE distributions.

The additional constraints provided by TTE data can theoretically mitigate the scaling and swapping ambiguity, but do not fully eliminate it \citep{hanin2014identifiability}. The guaranteed identifiability requires a sufficient follow-up window, meaning that the support of the diagnosis times among the positive lies within that of the censoring times \citep{maller1992immunes, li2001identifiability}. Empirically, this appears as a survival curve that eventually reaches a plateau, indicating individuals remaining undiagnosed at the end of follow-up are all truly negative \citep{amico2018cure}. When the follow-up is short and the plateau is poorly supported by negative cases, the estimated time model $f$ can be unreliable and distort classification \citep{othus2020bias, cortes2022msr137}.

\section{Simulations}\label{sec: simulation}

We conduct a comprehensive simulation study to evaluate the classification performance and robustness of MCMs with TTE modeling comparing to PU learning methods that do not have access to timing data. We design the simulation under a variety of challenging conditions, including varying sample size, outcome rate, censoring rate, and the underlying mechanisms of censoring and susceptibility.

We consider two methods that require SCAR assumption, namely Elkan-Noto PU (ENPU) \citep{elkan2008learning} and uPU \citep{du2015convex}, and two variants of SARPU \citep{bekker2019beyond} to represent different levels of feature constraint. The two SARPU variants differ in their prior knowledge regarding true propensity features: SARPU-F uses the full available feature set for both the probability and propensity models, while SARPU-R uses a restricted feature set for the propensity model that contains only the features truly relevant to propensity. Similarly, we analyze two variants of the Accelerated Failure Time Mixture Cure (AFTMC)  \citep{yamaguchi1992accelerated} to represent parametric MCMs, in which the TTE model is based on the full feature set (AFTMC-F) or a restricted set (AFTMC-R). We also include the discrete-time neural network (DTNN) as a more flexible neural network comparator. Two reference baselines are included: a logistic regression (LR) fit directly to the diagnosis indicator, and an AFT model that assumes universal susceptibility and uses predicted survival past maximum follow-up time as a outcome probability proxy. A logistic regression fit to the latent labels $Z$ serves as an oracle upper bound.

\subsection{Setup}

To simulate a right-censored population for binary classification tasks, we construct a data-generating process (DGP) that allows precise control over key elements. The DGP couples a feature-dependent outcome model with a feature-dependent diagnosis-time distribution for positive patients, subject to right-censoring. Each synthetic individual carries 4 covariates; outcome is governed by the first two through a logistic function, and diagnosis time by the remaining two through an AFT model, so that the two processes depend on disjoint feature sets. Censoring times are feature-independent and drawn from a fixed log-normal distribution tuned to the scale of the generated diagnosis times. In the baseline setup, we use this DGP to sample data for a total of $4000$ synthetic individuals and choose parameters to create a balanced sample with a 50\% outcome rate and a 50\% right-censoring rate among the positive individuals. Together with our baseline DGP, we vary the DGP across four key dimensions to evaluate each method's robustness and suitability across a range of challenging conditions, summarized here and specified fully in Appendix~\ref{app: dgp}.

\subsection{Results}\label{sec: results}
 
We use the area under the receiver operating characteristic curve (AUC-ROC) as the primary metric to evaluate classification performance for each method. Table~\ref{tab2} presents the average AUC and standard deviation.
 
\begin{figure*}[!t]
\floatconts
  {fig: ident}%
  {\caption{Analysis of model identifiability for AFTMC and SARPU across various scenarios. Each scatter point shows the AUC for predicted outcome probability and for predicted labeling propensity, both calculated against true positive labels. This quantifies the degree to which information about outcome is either (a) correctly captured by the outcome model (x-axis), or (b) incorrectly captured by the labeling propensity model (y-axis). An ideal model would exhibit high x-values and y-values around 0.5, except in the cases of positive or negative correlation (as illustrated in the SARPU-R).}}%
  {\includegraphics[width=0.85\linewidth]{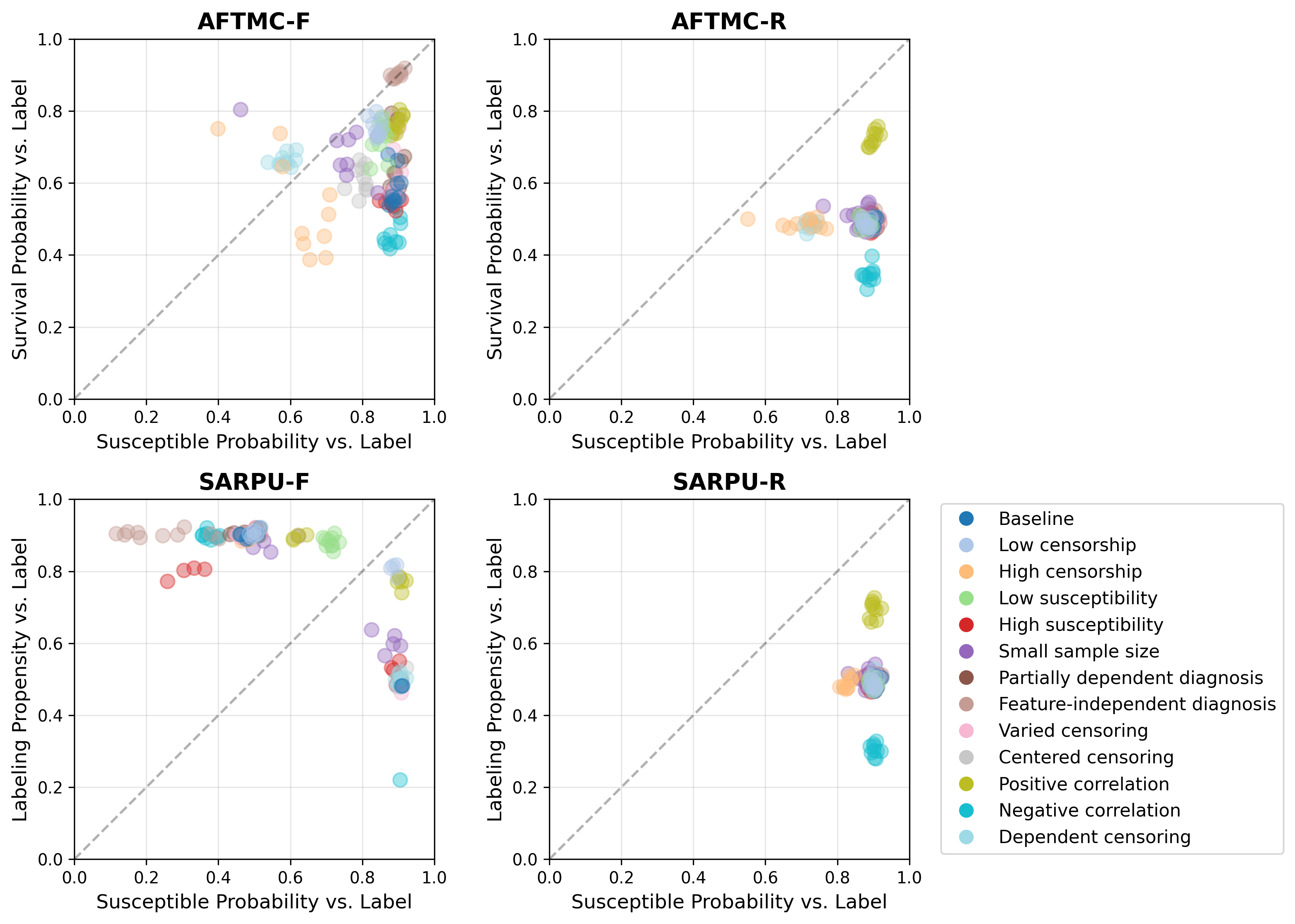}}
\end{figure*}

\begin{table*}[!t]
\floatconts
  {tab2}%
  {\caption{Average (standard deviation) AUC performance across 10 iterations for simulation experiments. Full simulation results are provided in Appendix~\ref{app: extra}.}}%
  {\resizebox{\textwidth}{!}{%
  \begin{threeparttable}
  \begin{tabular}{lrrrrrrr}
\toprule
 & \textbf{uPU} & \textbf{SARPU-F} & \textbf{SARPU-R} & \textbf{AFTMC-F} & \textbf{AFTMC-R} & \textbf{DTNN} & \textbf{LR} \\

\midrule
\multicolumn{8}{@{\hspace{0pt}}l}{\textbf{Exp.0: Baseline}} \\
Baseline & 0.83 (.02) & 0.57 (\underline{.18}) & \textbf{0.90 (.01)} & \textbf{0.89 (.01)} & \textbf{0.89 (.01)} & 0.83 (.01) & 0.81 (.02) \\

\midrule
\multicolumn{8}{@{\hspace{0pt}}l}{\textbf{Exp.1: Sample sizes and outcome rates}} \\
Low censorship & \textbf{0.89 (.01)} & 0.66 (\underline{.20}) & \textbf{0.90 (.01)} & 0.84 (.01) & 0.88 (.01) & \textbf{0.90 (.01)} & 0.88 (.01) \\
High censorship & 0.73 (.02) & 0.49 (.02) & \textbf{0.83 (.01)} & 0.63 (\underline{.10}) & 0.70 (.06) & 0.76 (\underline{.12}) & 0.78 (.02) \\
Low susceptibility & 0.87 (.01) & 0.71 (.01) & \textbf{0.90 (.01)} & 0.85 (.02) & 0.88 (.01) & \textbf{0.89 (.01)} & 0.86 (.01) \\
High susceptibility & 0.71 (.02) & 0.66 (\underline{.30}) & \textbf{0.89 (.01)} & 0.88 (.02) & \textbf{0.89 (.01)} & 0.73 (.06) & 0.67 (.02) \\
Small sample size & 0.81 (.03) & 0.71 (\underline{.18}) & \textbf{0.88 (.02)} & 0.75 (\underline{.11}) & 0.85 (.04) & 0.84 (.06) & 0.79 (.03) \\

\midrule
\multicolumn{8}{@{\hspace{0pt}}l}{\textbf{Exp.2: Feature-dependent propensity}} \\
Partially dependent diagnosis & 0.88 (.01) & 0.48 (.03) & \textbf{0.90 (.01)} & \textbf{0.89 (.01)} & \textbf{0.90 (.01)} & \textbf{0.90 (.01)} & 0.86 (.01) \\
Feature-independent diagnosis & \textbf{0.90 (.01)} & 0.24 (\underline{.10}) & \textbf{0.90 (.01)} & \textbf{0.90 (.01)} & \textbf{0.90 (.01)} & \textbf{0.90 (.01)} & \textbf{0.90 (.01)} \\
Varied censoring & 0.88 (.01) & 0.52 (\underline{.14}) & \textbf{0.90 (.01)} & \textbf{0.89 (.01)} & \textbf{0.90 (.01)} & \textbf{0.90 (.01)} & 0.86 (.01) \\
Centered censoring & 0.77 (.02) & 0.74 (\underline{.22}) & \textbf{0.90 (.01)} & 0.80 (.02) & \textbf{0.89 (.01)} & 0.78 (.05) & 0.77 (.02) \\

\midrule
\multicolumn{8}{@{\hspace{0pt}}l}{\textbf{Exp.3: Susceptibility-propensity correlation}} \\
Positive correlation & 0.87 (.01) & 0.79 (\underline{.15}) & \textbf{0.90 (.01)} & \textbf{0.90 (.01)} & \textbf{0.90 (.01)} & 0.87 (.01) & 0.86 (.01) \\
Negative correlation & 0.79 (.02) & 0.43 (\underline{.17}) & \textbf{0.90 (.01)} & 0.88 (.02) & \textbf{0.89 (.01)} & 0.85 (.01) & 0.75 (.02) \\

\midrule
\multicolumn{8}{@{\hspace{0pt}}l}{\textbf{Exp.4: Dependent censoring times}} \\
Dependent censoring & 0.72 (.02) & \textbf{0.90 (.01)} & \textbf{0.90 (.01)} & 0.59 (.02) & 0.73 (.01) & 0.85 (.03) & 0.75 (.02) \\
\bottomrule
\end{tabular}
\begin{tablenotes}[flushleft]
\item The best performances in each row are bolded. Standard deviations greater than 0.10 are underlined to highlight scenarios with higher performance variance. The oracle performance is $0.90$ across all experimental scenarios.
\end{tablenotes}
\end{threeparttable}%
}}
\end{table*}

\textbf{Exp.0 Baseline Scenario.} The baseline results show that methods that account for labeling selection bias substantially outperform SCAR-based methods, confirming the latter's strong reliance on the SCAR assumption. SARPU-R and both AFTMC variants achieve comparable performance approaching the oracle, while SARPU-F completely fails. This suggests that correctly identifying features that derive the selective labeling provides comparable restriction to what a TTE model could provide. Additionally, DTNN underperforms AFTMC-F despite sharing a similar objective function with AFTMC. We hypothesize that DTNN's high flexibility makes it prone to overfitting, introducing estimation errors that are inadvertently absorbed by the outcome probability model.

\textbf{Exp.1 Sample Sizes and Outcome Rates.} We then vary sample sizes and outcome rates to alter how much fully observed data is available and how contaminated the unlabeled set becomes. Not surprisingly, we find higher censoring rates and smaller sample sizes consistently reduce performance. Interestingly, raising the positive fraction cuts both ways: most methods degrade because the unlabeled set becomes more contaminated, whereas MCMs benefit from it since a larger positive population provides more information to model the TTE distribution, and therefore allows better predictions.

\textbf{Exp.2 Feature-Dependent Propensity.} This setting varies how strongly patient features drive labeling propensity. We control by either adjusting the signal-to-noise ratio in diagnosis times, or varying the spread of the stochastic censoring time distribution. The results from both settings show a consistent trend. As the diagnosis times become more stochastic, the gap between SCAR-based methods and the rest narrows. Once the diagnosis timing is fully feature-independent, almost all methods converge to an upper-bound AUC of $0.90$. This indicates that when censoring happens fully at random, the classification task becomes straightforward and complex constraints are no longer needed. Widening the censoring distribution similarly weakens the coupling between diagnosis time and propensity, steering the data toward the SCAR regime. Conversely, a tightly centered censoring time distribution forces individuals with late diagnosis times into the unlabeled set, exacerbating selection bias and causing SCAR-based methods to degrade significantly.

\textbf{Exp.3 Susceptibility-Propensity Correlation.} We investigate the effect of shared features that influence both an individual's outcome and their time to diagnosis. We first study the setting of positive correlation, where high-risk individuals are also more likely to be diagnosed early and labeled as positive. All methods improve in this setting because high-risk positives and low-risk negatives are cleanly separated in feature space, allowing models to easily learn the decision boundary. Conversely, negative correlation hurts most of them, where the most informative positive individuals are systematically censored out of the labeled set. Despite this difficulty, SARPU-R and AFTMC families still demonstrate good performance, showing the robustness of both propensity feature space constraint and TTE model constraint.

\textbf{Exp.4 Dependent Censoring Times.} Lastly, the hardest case is to make outcome probability, diagnosis time and censoring time are all covariate-driven. We find that AFTMC-F fails to model the time-to-diagnosis distribution when irrelevant censoring-related features are included, likely because its rigid parametric structure cannot disentangle the effects of censoring-related variation from the latent outcome probability and event-time components. In contrast, the DTNN and SARPU-R demonstrate superior robustness. The DTNN learns a flexible, discrete-time model of the distribution of the observed times of diagnosis or censoring events, allowing it to learn that censoring-related features affect only observation times rather than the outcome probability. Similarly, SARPU-R succeeds by avoiding the rigid AFT structures entirely, using a generalized propensity term that allows the model to absorb the complexity from the dependent censoring without corrupting the estimation of the true outcome probability.

\subsection{Discussion of Model Identifiability}\label{sec: ident}

The superior performance of SARPU-R, AFTMC, and DTNN highlights the advantage of methods that can model both the outcome probability and labeling propensity. However, a major challenge for these models is identifiability. Figure~\ref{fig: ident} plots the AUC for the predicted outcome probability and the predicted labeling propensity against true labels from AFTMC and SARPU, showing how outcome probability information is captured by each model component. The results show that without proper constraints, these models can conflate the outcome probability and the labeling propensity, leading to unreliable predictions. This is particularly evident in the unconstrained SARPU-F, where the true outcome probability is instead captured by the labeling propensity model. This can be attributed to a systematic bias in the SARPU optimization process: the propensity model learns directly from the censoring indicators, quickly converging while identifying relationships between features and the probability of censoring. The probability model, however, left with a more ambiguous task, is then prone to being ``permuted'' with the propensity model, as the model defaults to solving the easier problem first. In contrast, methods with an explicit constraint on propensity features, like AFTMC-R and SARPU-R, successfully separate the two components, demonstrating the critical importance of external information to achieve robust performance. 

The inherent TTE structure in AFTMC provides an effective natural constraint that helps mitigate the identifiability issue. However, this time-based constraint is not perfect; it allows for minor interference between the components, forcing the probability model to compensate for minor inaccuracies in the propensity model and resulting in a less accurate representation of the true underlying outcome. Despite that, TTE modeling is still a robust and preferred constraint when prior knowledge about propensity features is not available.

\section{Real-World Experiments}\label{sec: real}

In this section, we compare different methods on two clinical datasets to evaluate their prediction performance as well as their robustness in real-world applications. To better understand the underlying censoring mechanism and evaluate against ground-truth labels, we first investigate a semi-synthetic diabetes dataset with simulated diagnoses and censoring times. We then analyze an autism screening study under real, unmanipulated censoring scenarios.

\subsection{Diabetes Study with Synthetic Censoring}

We resample the CDC diabetes incidence study \citep{burrows2017incidence} into a balanced cohort of $10{,}000$ patients, preserving real demographics, lab results and ground-truth diagnosis labels, and then mask a portion of the diagnoses using synthetic diagnosis and censoring times generated as in Section~\ref{sec: simulation}. The full experiment details are provided in Appendix~\ref{app: diabetes} with results in Table~\ref{tb: diabetes}.

The pattern observed in this semi-synthetic experiment is consistent with our previous full simulation settings. SARPU-R and AFTMC-R, which restrict the propensity model to the features that truly govern labeling, consistently perform best, and SARPU-R attains the best or tied-best AUC in all seven scenarios at $0.82$ without using any diagnosis times. Where such prior knowledge is withheld, AFTMC-F remains the strongest comparator, as the TTE component effectively helps in understanding the underlying censoring process.

\begin{figure}[!t]
\floatconts
  {fig:asd}%
  {\caption{The boxplots show the AUC scores for different methods used in the autism screening study. All methods are evaluated across 5 random seeds. AP results are included in Appendix~\ref{app: asd}.}}%
  {\includegraphics[width=\linewidth]{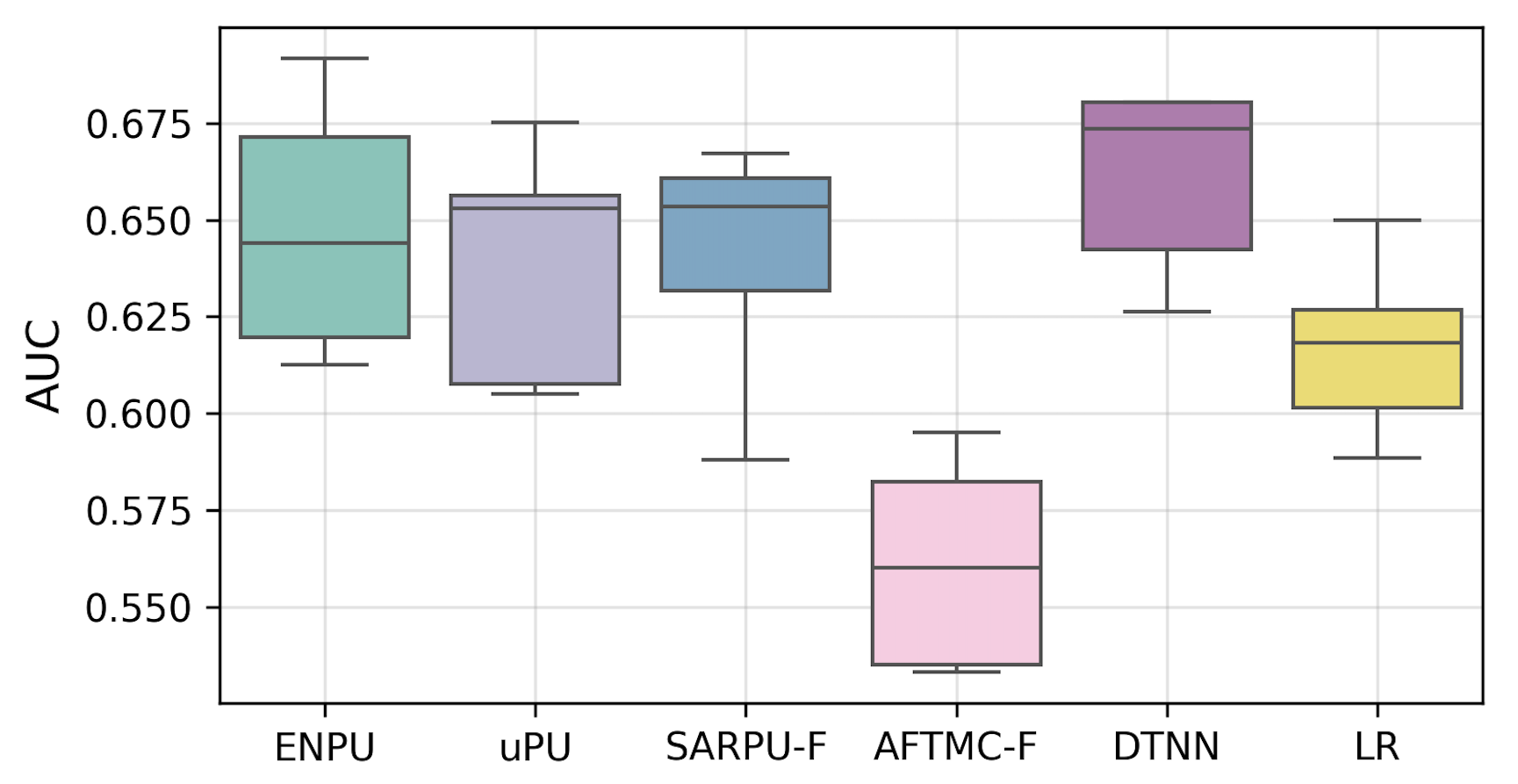}}
\end{figure}

\subsection{Autism Screening Study}

We then analyze an autism screening dataset containing real right-censored patient trajectories. The data is extracted from the Clinical Research Data Mart (CRDM) at Duke University \citep{hurst2021development}. Each child is represented by a $256$-dimensional vector averaging word2vec medical concept embeddings over all electronic health record (EHR) codes recorded before age two, concatenated with demographics. We impose an artificial study endpoint that applies uniform administrative censoring across the cohort to emulate a prospective setting, yielding $33{,}137$ children of whom only $1.28\%$ are diagnosed. The censoring remains realistic since the censoring times derive from the original trajectories. For evaluation, we retain only test-set children with follow-up beyond age nine and treat those without a recorded diagnosis as ground-truth negative. This is the one setting in which a feature restriction is genuinely unavailable: the abstract and high-dimensional feature space makes it infeasible to name the subset that drives time to diagnosis; therefore, additional constraints such as TTE modeling become necessary.

Figure~\ref{fig:asd} shows that the DTNN achieves the best overall performance. Its flexible TTE structure is well-suited for such large datasets and ensures easy convergence even in complex real-world censoring scenarios. AFTMC-F, which imposes the same restriction through a rigid parametric form, instead performs poorly in high-dimensional feature space. In the absence of TTE data, which in some cases may be unreliable or costly to obtain, PU learning methods serve as good alternatives, with all tested PU approaches significantly outperforming the LR baseline.

\section{Conclusions}\label{sec: con}

In this study, we propose a unified framework for binary classification with right-censored clinical data, casting all method families as a single likelihood optimization problem under different restrictions on the labeling propensity. Within this framework, an MCM is essentially a SAR-PU model whose propensity is parameterized through the diagnosis and censoring time distributions rather than estimated freely. The TTE component therefore functions as a restriction on the propensity rather than an independent source of information about outcome, which reframes the question of whether it is necessary as a question of whether TTE restriction is the best available.

Our answer is that the TTE component earns its cost in one regime only, and the practical guidance follows the order in which restrictions become available. We summarize the conditions and corresponding beneficial methods into a decision flowchart in Figure~\ref{fig: Reccomend} in Appendix~\ref{app: flow}. 

We briefly summarize the decision logic as follows. If diagnosis times are feature-independent, or censoring times are broadly dispersed relative to them, no restriction is needed at all. If censoring happens selectively but the features governing it can be identified \textit{a priori}, a SCAR-free PU model restricted to those features suffices. Only when the TTE features cannot be explicitly named does the TTE component become worthwhile, as in a high-dimensional EHR cohort or complex clinical scenarios.

\acks{This work was supported by the U.S. National Institute of Mental Health
(grant no.\ K01 MH127309).}

\bibliography{ref}

\appendix

\section{Related Work}\label{app: review}

\subsection{Mixture Cure Models}\label{app: mcm}

Extensive research on MCMs has led to a variety of implementations,
which differ in how strongly the event-time distribution is
parameterized. The classification model is commonly a logistic
regression, while the TTE model can take various forms. Fully
parametric approaches specify the baseline event-time distribution,
for instance Exponential \citep{farewell1977model} or Weibull
\citep{farewell1982use}, with covariate effects usually modeled
under an accelerated failure time (AFT) framework in which
covariates act multiplicatively on the time scale
\citep{yamaguchi1992accelerated}. Semi-parametric methods instead
leave the baseline unspecified, including the Cox proportional
hazards formulation \citep{kuk1992mixture} and semi-parametric AFT
extensions \citep{li2002semi}. More recent developments include
nonparametric methods that estimate both components using
kernel-based or smoothing spline techniques
\citep{lopez2017nonparametric, patilea2020general}.

The Discrete-Time Neural Network (DTNN) sits at the flexible end
of this progression. It discretizes the continuous time axis into
a sequence of intervals (``time bins'') and predicts the
probability of event occurrence within each, with a final bin
representing occurrence beyond the time horizon, which corresponds
to the cure probability in the MCM setting \citep{yu2011learning}.
The neural formulation scales to large datasets and the
discrete-time parameterization is numerically stable.

The effectiveness of MCMs in outcome classification depends heavily on
correct specification of the TTE model \citep{hanin2014identifiability}. The
key difficulty is separating truly non-susceptible individuals from those
who are censored but still at risk. In theory, identifiability of the cure
fraction is guaranteed only when the survival curve eventually reaches a
plateau, indicating that the individuals remaining event-free are all truly
non-susceptible \citep{maller1992immunes, maller1994sufficient,
li2001identifiability, amico2018cure}. In practice, if the plateau is brief
and poorly supported by non-susceptible cases, follow-up has been too short
to distinguish non-susceptible individuals reliably, the estimated survival
function becomes unstable, and classification is distorted
\citep{othus2020bias, cortes2022msr137}. A further limitation is the strong
reliance on accurate and complete timing information; when event times are
missing or unreliable, the applicability of MCMs is severely compromised
\citep{cipriani2025missing, schulz2021temporal}.

\subsection{Positive-Unlabeled Learning}

While MCMs depend on event-time data, PU learning provides an alternative
approach to the same task without requiring timing information. Traditional
PU learning methods are mostly built on the SCAR assumption, under which the
labeling propensity is feature-independent \citep{elkan2008learning}. This
assumption simplifies the problem by allowing the class prior, the
proportion of susceptibles in the unlabeled set, to be estimated. Early
approaches such as that of \citet{elkan2008learning} use a two-step process,
first estimating the labeling frequency from the unlabeled set and then
training a standard classifier using this estimate. More recent methods
improve on this by developing direct, unbiased empirical risk estimators:
unbiased PU (uPU) \citep{du2014analysis, du2015convex} defines an unbiased
training objective, while non-negative PU \citep{kiryo2017positive} further
stabilizes training by keeping the estimated risk non-negative.

The SCAR assumption is rarely realistic in clinical settings, and
particularly not under right-censoring, where diagnosis times are typically
feature-dependent so that patients diagnosed earlier have a higher
propensity to be labeled. Recent advances in PU learning have focused on
relaxing SCAR, producing methods that closely resemble the two-component
structure of MCMs: a binary classifier estimating susceptibility probability
is paired with a separate model for the labeling propensity
\citep{bekker2019beyond, gong2021instance}. This design lets PU learning
handle the censoring mechanism more directly and flexibly than a parametric
MCM.

Even so, PU learning faces a fundamental challenge: without modeling the
event-time distribution, susceptibility probability and labeling propensity
are hard to separate, especially when the features driving the two processes
are correlated. PU learning therefore relies on additional assumptions to
disentangle the components \citep{bekker2020learning}.
\citet{bekker2019beyond} assume that the propensity depends only on a
restricted subset of the observed features while the susceptibility model
uses the full set. \citet{kato2019learning} assume an ``invariance of
order'' property, under which the likelihood of observing a label aligns
with the latent risk. \citet{gong2021instance} impose the more restrictive
condition that all censored instances have propensities below a fixed
threshold. \citet{gerych2022recovering} show that the propensity is
identified either by restricting its functional form or by assumptions that
avoid doing so, which is the dichotomy Section \ref{sec: relation} builds
on. These assumptions are strong and context-dependent, and must be applied
with care in practice.

\section{Method Details}\label{app: methods}

\subsection{SCAR-Based PU Learning}
\label{app: scar}

This appendix gives the two SCAR-based objectives summarized in
Section~\ref{sec: scar} and their derivation from the
complete-data log-likelihood of
Equation~\ref{eq: complete_lld}.

Under SCAR the propensity is fixed to a constant, $e(\X)=c$, and
Equation~\ref{eq: complete_lld} simplifies considerably. Because
$S=1$ implies $Z=1$, we have $s_iz_i=s_i$ and
$(1-s_i)(1-z_i)=1-z_i$, so the likelihood separates into a term
involving only $p$,
\begin{equation}\label{eq: oracle_lld}
    \begin{aligned}
    l_{\mathrm{orc}}(p;z) = \sum_{i=1}^n
      & z_i\ln[p(\x_i)] \\
      &+ (1-z_i)\ln[1-p(\x_i)],
    \end{aligned}
\end{equation}
and a term involving only $c$, namely
$\sum_i s_i\ln c + (1-s_i)z_i\ln(1-c)$. Fixing the value of $e$
thus decouples the two components entirely:
Equation~\ref{eq: oracle_lld} is the objective that would be
optimized if the latent labels were observed, and the whole
difficulty reduces to recovering it from the $s_i$. The two
SCAR-based methods correspond to the two ways of doing so.

\citet{elkan2008learning} eliminate $Z$ by marginalization.
Summing the complete-data likelihood over $z_i\in\{0,1\}$ with
$e\equiv c$ gives $\Pr(S=1\mid\x)=c\,p(\x)$ and
\begin{equation*}
    \begin{aligned}
    \Pr(S=0\mid\x) &= p(\x)(1-c) + 1-p(\x) \\
                   &= 1-c\,p(\x),
    \end{aligned}
\end{equation*}
where the two summands on the first line are the censored
susceptible and the non-susceptible explanations of $s_i=0$. The
resulting marginal log-likelihood over $Z$ is
\begin{equation}\label{appeq: marginal_lld_s}
    \begin{aligned}
    l(p;\mathcal{D}_{binary},c) = \sum_{i=1}^n
      & s_i\ln[c\cdot p(\x_i)] \\
      &+ (1-s_i)\ln[1-c\cdot p(\x_i)],
    \end{aligned}
\end{equation}
with $c$ entering as a fixed scaling parameter. It is estimated by
training a classifier for $\Pr(S=1\mid\X)$ and averaging its
predictions over the labeled susceptibles, which is valid because
SCAR renders them representative of all susceptibles, so that
$\mathbb{E}_{S=1}\Pr(S=1\mid\X)=\mathbb{E}_{Z=1}\Pr(S=1\mid\X)=c$.

\citet{du2014analysis} instead retain
Equation~\ref{eq: oracle_lld} and replace $z_i$ by an unbiased
surrogate, treating every unlabeled case as a proxy for the
non-susceptible and correcting for the resulting inflation of
non-susceptibles and underestimation of susceptibles. Under SCAR,
$\mathbb{E}[S\mid\X]=c\,p(\X)$, so $S/c$ and $1-S/c$ are unbiased
substitutes for $Z$ and $1-Z$ in any expectation of the form
$\mathbb{E}[Z\,h(\X)]$. Substituting them into
Equation~\ref{eq: oracle_lld} yields the unbiased PU (uPU)
objective,
\begin{equation}\label{eq: marginal_lld_z}
    \begin{aligned}
    \hat{l}(p;\mathcal{D}_{binary},c) = \sum_{i=1}^n
      & \tfrac{s_i}{c}\ln[p(\x_i)] \\
      &+ \Big(1-\tfrac{s_i}{c}\Big)\ln[1-p(\x_i)],
    \end{aligned}
\end{equation}
in which labeled and unlabeled instances are weighted so that the
empirical loss remains unbiased. This objective is commonly paired
with robust classification losses such as the convex
\citep{du2015convex} and non-negative \citep{kiryo2017positive}
losses; we use the log-likelihood loss to align with our unified
framework.

The two routes differ in where the constraint is absorbed.
Equation~\ref{appeq: marginal_lld_s} marginalizes $Z$ out and keeps
all weights in $[0,1]$, at the cost of depending on the estimated
$c$ through a plug-in. Equation~\ref{eq: marginal_lld_z} keeps
the oracle objective and moves $c$ into the weights, which holds
in expectation but not pointwise: the coefficient $1-s_i/c$ is
negative on labeled instances, so the objective is an unbiased
risk estimator rather than a log-likelihood, and a sufficiently
flexible model can drive it below the attainable minimum. This is
the failure mode that non-negative PU \citep{kiryo2017positive}
corrects by clipping the unlabeled-class risk at zero.

Both ENPU and uPU are implemented with a neural network classifier
optimized under a negative log-likelihood loss, so that they
differ from the SCAR-free comparators only in the restriction
imposed on $e$ and not in the hypothesis class.
\subsection{SARPU Expectation--Maximization}

Rather than maximizing the marginal log-likelihood of Equation
\ref{eq: marginal_lld_e} directly, SARPU applies EM to the complete-data
log-likelihood of Equation \ref{eq: complete_lld}, which is numerically more
stable. With $\hat p$ and $\hat e$ the current iterates, the E-step imputes
the latent susceptibility as
$$\hat{z}_i = s_i + (1-s_i)\frac{\hat{p}(\x_i)(1-\hat{e}(\x_i))}{1-\hat{p}(\x_i)\hat{e}(\x_i)},$$
and the M-step updates the two components separately,
$$\operatorname*{argmax}_{\hat{p}}\sum_{i=1}^n\big[\hat{z}_i\ln \hat{p}(\x_i) +(1-\hat{z}_i)\ln(1-\hat{p}(\x_i))\big],$$
$$\operatorname*{argmax}_{\hat{e}}\sum_{i=1}^n\hat{z}_i\big[s_i\ln \hat{e}(\x_i)+(1-s_i)\ln(1-\hat{e}(\x_i))\big].$$
The second display makes the asymmetry noted in Section \ref{sec: ident}
explicit: the propensity update is a supervised regression of $s_i$ on
$\x_i$ among the imputed susceptibles, and therefore converges quickly,
whereas the susceptibility update depends on the imputed $\hat z_i$ and is
correspondingly slower to settle. When no feature restriction separates the
two, this is the mechanism by which the components swap.

\subsection{Optimizing Mixture Cure Models}

MCMs may be fit by the same EM strategy, with the propensity $\hat e(\x_i)$
in the E-step replaced by the survival term $1-\hat F(\x_i)$ and the M-step
updating $\hat p$ and $\hat f$ jointly. The TTE component, however, makes
the expected complete-data log-likelihood strongly non-convex, so EM becomes
sensitive to initialization and converges slowly. We therefore optimize the
marginal log-likelihood of Equation \ref{eq: marginal_lld_mcm} directly by
gradient descent, which is computationally cheaper and, in our experiments,
more stable. AFTMC is composed of a logistic susceptibility component and an
AFT event-time component, each instantiated as a neural network. The DTNN is
defined and trained following \citet{engelhard2022disentangling}, with time
partitioned into 10 intervals of approximately equal event counts.

\section{Simulation Design}\label{app: dgp}

This appendix gives the data-generating process and the parameter settings
summarized in Section \ref{sec: simulation}.

\subsection{Baseline Data-Generating Process}

In the baseline setup (Exp. 0), we generate a set of synthetic individuals, each characterized by four random covariates $\x=[x_1, x_2, x_3, x_4]$. The susceptibility probability $p(\x)$ is modeled by the first two covariates through a logistic function $p(\x) = \sigma(2x_1-2x_2)$, and the true, latent susceptibility status ($Z$) is then sampled from a Bernoulli distribution with probability $p(\x)$. The time-to-event distribution follows an AFT model, where the diagnosis times are determined by the remaining covariates $t= \exp(2x_3-2x_4+\epsilon)$, with the noise term $\epsilon\sim \mathcal{N}(0, \sigma_\epsilon^2)$. We use the variance $\sigma^2_\epsilon$ to control how strongly the diagnosis times depend on the covariates, and set $\sigma_\epsilon^2=0$ in the baseline scenario to isolate the effect of feature-dependent time-to-diagnosis. The censoring times are feature-independent and randomly sampled from a fixed log-normal distribution, which is tuned to match the scale and variability of the generated diagnosis times.

\subsection{Variation Dimensions}

\begin{itemize}
    \item \textbf{E1. Sample sizes and outcome rates.} These scenarios manipulate the basic aspects of the training data, including the censoring rate, susceptibility rate, and sample size. We adjust the censoring rate by scaling our synthetic censoring times, and we modify the susceptibility rate by tuning the intercept in the logistic model. These adjustments directly alter the number of true susceptible and true non-susceptible instances, as well as the balance of labeled and unlabeled data.
    \item \textbf{E2. Feature-dependent propensity.}
    These scenarios vary the degree to which individuals' features influence the labeling propensity by controlling the effect of features on diagnosis times or censoring times. We first vary the degree of feature-dependence by adjusting the signal-to-noise ratio when generating diagnosis times while preserving the marginal distribution. When diagnosis times are fully driven by random noise, the labeling mechanism becomes independent of individuals' features, thus conforming to the SCAR assumption. We also consider different censoring patterns by controlling the variance of the censoring time distribution. A higher-variance censoring distribution represents loss to follow-up over a wide range of times, reducing the sensitivity of the labeling propensity to feature-dependent diagnosis times. In contrast, a lower-variance, more deterministic censoring process concentrates censoring times within smaller time intervals, amplifying the effect of diagnosis-time feature-dependence on the labeling propensity.
    \item \textbf{E3. Susceptibility--propensity correlation.} These scenarios explore how the correlation between susceptibility probability and labeling propensity, arising from their respective dependence on shared features, affects performance. In the \textit{positive correlation} scenario, high-risk individuals are diagnosed early and thus are more likely to be labeled as susceptible. Conversely, in the \textit{negative correlation} scenario, high-risk individuals are more likely to be censored, systematically removing individuals most likely to be susceptible to diagnosis from the labeled set.
    \item \textbf{E4. Dependent censoring times.} This scenario considers the most complex condition of dependent censoring, where susceptibility probability, time-to-diagnosis, and time-to-censoring are all directly influenced by covariates.
\end{itemize}

The specifications for each of these experimental variants are provided in Table \ref{tab1}.

\begin{table*}[!t]
\floatconts
  {tab1}%
  {\caption{Simulation Study Design and Experimental Variants.}}%
  {\small
  \begin{threeparttable}
  \begin{tabular*}{\textwidth}{@{\extracolsep\fill}lll@{\extracolsep\fill}}
\toprule
\textbf{Experiment} & \textbf{Scenario Name} & \textbf{Modified Parameters} \\
\midrule
Baseline & Baseline & Sample size = 4000\\
& & Censoring rate = 50\% \\
& & Susceptibility rate = 50\%\\
& & $z \sim Bernoulli(\sigma(2x_1-2x_2))$\\
& & $\log t = 2x_3-2x_4$, $\log c \sim N(0, \sigma_t^2)$\tnote{$\dagger$}\\
\midrule
Sample sizes and outcome rates & Low censorship & Censoring rate = 20\% \\
 & High censorship & Censoring rate = 80\% \\
 & Low susceptibility & Susceptibility rate = 20\% \\
 & High susceptibility & Susceptibility rate = 80\% \\
 & Small sample size & Sample size = 1000 \\
\midrule
Feature-dependent propensity & Partially dependent diagnosis & $\log t \sim N(x_3-x_4, 0.75\sigma_t^2)$ \\
 & Feature-independent diagnosis & $\log t \sim N(0,\sigma_t^2)$ \\
 & Varied censoring & $\log c \sim N(0,25\sigma_t^2)$ \tnote{$\ddagger$} \\
 & Centered censoring & $\log c \sim N(0,0.25\sigma_t^2)$ \\
\midrule
Susceptibility--propensity correlation  & Positive correlation & $\log t = -2x_1+2x_2$ \\
 & Negative correlation & $\log t = 2x_1-2x_2$ \\
\midrule
Dependent censoring times & Dependent censoring & $\log t = 2x_3, \log c =-2 x_4$ \\
\bottomrule
\end{tabular*}
\begin{tablenotes}
\item[$\dagger$] $\sigma_t$ is the standard deviation for $\log(t)$ in baseline setting.
\item[$\ddagger$] Censoring times $C$ are sampled from a truncated normal distribution to keep them within a reasonable range. When the variance is large, this distribution is nearly uniform.
\item Scenario names match those used in Table \ref{tab2}.
\end{tablenotes}
\end{threeparttable}}
\end{table*}

\section{Additional Simulation Results}\label{app: extra}

Table \ref{tab2} in the main text omits ENPU, which imposes the same
constant-propensity restriction as uPU and behaves almost identically to it,
and the AFT baseline, which assumes universal susceptibility. Table
\ref{tab2full} reports all nine comparators across all 13 scenarios. The
oracle logistic regression fit to the latent labels $Z$ attains an AUC of
$0.90$ in every scenario and is therefore not tabulated.

\begin{table*}[!t]
\floatconts
  {tab2full}%
  {\caption{Average (standard deviation) AUC across 10 iterations for all nine comparators in the simulation experiments.}}%
  {\resizebox{\textwidth}{!}{%
  \begin{threeparttable}
  \begin{tabular}{lrrrrrrrrr}
\toprule
 & \textbf{ENPU} & \textbf{uPU} & \textbf{SARPU-F} & \textbf{SARPU-R} & \textbf{AFTMC-F} & \textbf{AFTMC-R} & \textbf{DTNN} & \textbf{AFT} & \textbf{LR} \\
\midrule
\multicolumn{10}{@{\hspace{0pt}}l}{\textbf{Exp. 0: Baseline}} \\
Baseline & 0.81 (.02) & 0.83 (.02) & 0.57 (\underline{.18}) & \textbf{0.90 (.01)} & \textbf{0.89 (.01)} & \textbf{0.89 (.01)} & 0.83 (.01) & 0.79 (.02) & 0.81 (.02) \\
\midrule
\multicolumn{10}{@{\hspace{0pt}}l}{\textbf{Exp. 1: Sample sizes and outcome rates}} \\
Low censorship & 0.88 (.01) & \textbf{0.89 (.01)} & 0.66 (\underline{.20}) & \textbf{0.90 (.01)} & 0.84 (.01) & 0.88 (.01) & \textbf{0.90 (.01)} & 0.84 (.02) & 0.88 (.01) \\
High censorship & 0.71 (.02) & 0.73 (.02) & 0.49 (.02) & \textbf{0.83 (.01)} & 0.63 (\underline{.10}) & 0.70 (.06) & 0.76 (\underline{.12}) & 0.70 (.02) & 0.78 (.02) \\
Low susceptibility & 0.86 (.01) & 0.87 (.01) & 0.71 (.01) & \textbf{0.90 (.01)} & 0.85 (.02) & 0.88 (.01) & \textbf{0.89 (.01)} & 0.80 (.01) & 0.86 (.01) \\
High susceptibility & 0.67 (.02) & 0.71 (.02) & 0.66 (\underline{.30}) & \textbf{0.89 (.01)} & 0.88 (.02) & \textbf{0.89 (.01)} & 0.73 (.06) & 0.65 (.02) & 0.67 (.02) \\
Small sample size & 0.79 (.03) & 0.81 (.03) & 0.71 (\underline{.18}) & \textbf{0.88 (.02)} & 0.75 (\underline{.11}) & 0.85 (.04) & 0.84 (.06) & 0.76 (.02) & 0.79 (.03) \\
\midrule
\multicolumn{10}{@{\hspace{0pt}}l}{\textbf{Exp. 2: Feature-dependent propensity}} \\
Partially dependent diagnosis & 0.86 (.01) & 0.88 (.01) & 0.48 (.03) & \textbf{0.90 (.01)} & \textbf{0.89 (.01)} & \textbf{0.90 (.01)} & \textbf{0.90 (.01)} & 0.85 (.02) & 0.86 (.01) \\
Feature-independent diagnosis & \textbf{0.90 (.01)} & \textbf{0.90 (.01)} & 0.24 (\underline{.10}) & \textbf{0.90 (.01)} & \textbf{0.90 (.01)} & \textbf{0.90 (.01)} & \textbf{0.90 (.01)} & \textbf{0.90 (.01)} & \textbf{0.90 (.01)} \\
Varied censoring & 0.86 (.01) & 0.88 (.01) & 0.52 (\underline{.14}) & \textbf{0.90 (.01)} & \textbf{0.89 (.01)} & \textbf{0.90 (.01)} & \textbf{0.90 (.01)} & 0.83 (.02) & 0.86 (.01) \\
Centered censoring & 0.77 (.02) & 0.77 (.02) & 0.74 (\underline{.22}) & \textbf{0.90 (.01)} & 0.80 (.02) & \textbf{0.89 (.01)} & 0.78 (.05) & 0.71 (.02) & 0.77 (.02) \\
\midrule
\multicolumn{10}{@{\hspace{0pt}}l}{\textbf{Exp. 3: Susceptibility--propensity correlation}} \\
Positive correlation & 0.86 (.01) & 0.87 (.01) & 0.79 (\underline{.15}) & \textbf{0.90 (.01)} & \textbf{0.90 (.01)} & \textbf{0.90 (.01)} & 0.87 (.01) & 0.85 (.02) & 0.86 (.01) \\
Negative correlation & 0.75 (.02) & 0.79 (.02) & 0.43 (\underline{.17}) & \textbf{0.90 (.01)} & 0.88 (.02) & \textbf{0.89 (.01)} & 0.85 (.01) & 0.70 (.03) & 0.75 (.02) \\
\midrule
\multicolumn{10}{@{\hspace{0pt}}l}{\textbf{Exp. 4: Dependent censoring times}} \\
Dependent censoring & 0.75 (.02) & 0.72 (.02) & \textbf{0.90 (.01)} & \textbf{0.90 (.01)} & 0.59 (.02) & 0.73 (.01) & 0.85 (.03) & 0.68 (.03) & 0.75 (.02) \\
\bottomrule
\end{tabular}
\begin{tablenotes}[flushleft]
\item The best performances in each row are bolded. Standard deviations greater than 0.10 are underlined to highlight scenarios with higher performance variance. The oracle performance is $0.90$ across all experimental scenarios.
\end{tablenotes}
\end{threeparttable}%
}}
\end{table*}

\section{Diabetes Study Details}\label{app: diabetes}

\subsection{Cohort Construction and Synthetic Times}

We use a semi-synthetic dataset to evaluate the effectiveness of our comparator methods for identifying individuals susceptible to diabetes. This dataset is constructed by resampling the original diabetes data from the CDC's diabetes incidence study \citep{burrows2017incidence, CDC_Diabetes_2023} to obtain a balanced downstream cohort of $10{,}000$ patients. We preserve real-world patient characteristics, including demographics, lab results, and ground-truth diabetes diagnosis labels.

To simulate the challenge of learning an accurate diagnosis susceptibility model from right-censored outcomes, we introduce synthetic diagnosis and censoring times to mask a portion of diabetes diagnoses. The features used for generating synthetic diagnosis times are chosen to achieve a desired correlation between the labeling propensity and the susceptibility probability. Specifically, we first apply logistic regression to the original dataset to identify features with strong or weak correlation to diabetes risk. The most strongly correlated features, BMI and age, are incorporated into the time-to-diagnosis model to simulate scenarios where censoring is highly correlated with susceptibility probability, whereas the weakest correlates (income and education) are used to model scenarios where censoring propensity is largely uncorrelated with the susceptibility probability. The detailed process for generating these synthetic times is the same as in the simulation experiments of Section \ref{sec: simulation}, and the seven scenarios reported here correspond to the baseline, the four feature-dependent propensity variants and the two correlation variants of Table \ref{tab1}.

\subsection{Full Results}

Table \ref{tab3} reports all nine comparators across the seven scenarios.
The results are consistent with the simulation study. AFTMC-R and SARPU-R,
which account for the selective censoring process and restrict the
propensity model to the truly propensity-relevant features, consistently
achieve the best performance, and SARPU-R is slightly more robust than
AFTMC-R, attaining the best or tied-best AUC in all seven scenarios at
$0.82$ without using any diagnosis times. This suggests that dispensing with
the TTE component and modeling the censoring mechanism through a generalized
propensity model offers greater stability and predictive accuracy when the
relevant features are known. When such prior knowledge is unavailable,
AFTMC-F remains the most effective method, as it uses the timing information
to model the underlying censoring process; the unrestricted SARPU-F is the
weakest throughout.

\begin{table*}[!t]
\floatconts
  {tab3}%
  {\caption{Average (standard deviation) AUC across 10 iterations for the diabetes study with synthetic censoring.}}%
  {\resizebox{\textwidth}{!}{%
  \begin{threeparttable}
  \begin{tabular}{lrrrrrrrrr}
\toprule
\textbf{Scenario} & \textbf{ENPU} & \textbf{uPU} & \textbf{SARPU-F} & \textbf{SARPU-R} & \textbf{AFTMC-F} & \textbf{AFTMC-R} & \textbf{DTNN} & \textbf{AFT} & \textbf{LR} \\
\midrule
Baseline & 0.72 (.01) & 0.71 (.01)  & 0.54 (.05) & \textbf{0.82 (.01)} & 0.79 (.01) & \textbf{0.81 (.01)} & 0.65 (.04) & 0.68 (.01) & 0.72 (.01) \\
\midrule
Partially dependent diagnosis & 0.77 (.01) & 0.77 (.01) & 0.55 (.06) & \textbf{0.82 (.01)} & 0.80 (.01) & \textbf{0.81 (.01)} & 0.77 (.02) & 0.75 (.01) & 0.77 (.01) \\
Feature-independent diagnosis & 0.76 (.01) & 0.76 (.01) & 0.65 (\underline{.13}) & \textbf{0.82 (.01)} & 0.68 (.04) & 0.76 (.01) & 0.80 (.01) & 0.71 (.01) & 0.77 (.02) \\
Varied censoring & 0.78 (.01) & 0.79 (.01) & 0.60 (.07) & \textbf{0.82 (.01)} & \textbf{0.81 (.01)} & \textbf{0.81 (.01)} & \textbf{0.81 (.01)} & 0.75 (.01) & 0.78 (.01) \\
Centered censoring & 0.69 (.01) & 0.70 (.01) & 0.62 (.08) & \textbf{0.82 (.01)} & 0.67 (.03) & 0.77 (.02) & 0.66 (.06) & 0.63 (.01) & 0.69 (.01) \\
\midrule
Positive correlation & 0.80 (.02) & \textbf{0.80 (.01)} & 0.61 (.06) & \textbf{0.80 (.01)} & 0.79 (.01) & \textbf{0.80 (.01)} & 0.32 (.02) & 0.79 (.02) & \textbf{0.80 (.01)} \\
Negative correlation & 0.66 (.01) & 0.72 (.02) & 0.57 (\underline{.11}) & \textbf{0.82 (.01)} & 0.78 (.01) & 0.79 (.01) & 0.78 (.02) & 0.57 (.02) & 0.67 (.03) \\
\bottomrule
\end{tabular}\label{tb: diabetes}
\begin{tablenotes}[flushleft]
\item The best performances in each row are bolded. Standard deviations greater than 0.10 are underlined.
\end{tablenotes}
\end{threeparttable}%
}}
\end{table*}

\section{Autism Cohort Details}\label{app: asd}

The autism screening data is extracted from the Clinical Research Data Mart (CRDM)
at Duke University \citep{hurst2021development}. The cohort includes $55{,}229$
children born in 2014--2023 who had a well-child visit between 12 and 24
months of age and at least one follow-up visit after 24 months. For each
child, we construct a 256-dimensional feature vector by averaging
word2vec-based medical concept embeddings of all electronic health record
(EHR) codes recorded before age 2, then concatenate the demographic
variables to form the final patient representation. The full observation
period extends through June 2025, when the oldest children had reached 11
years of age and $3.42\%$ had received an autism diagnosis.

To emulate a realistic prospective setting and directly investigate
prediction under right-censoring, we establish an earlier, artificial study
endpoint at December 2022. This imposes uniform administrative censoring
across the cohort which, unlike individual-specific loss to follow-up,
better aligns the resulting data with the non-informative censoring
assumption. By masking any autism diagnoses and truncating trajectories
beyond this date, the procedure exaggerates the gap between the observed
diagnosis indicators available in training and the true underlying
susceptibility labels. The result is a heavily censored dataset in which the
observed autism diagnosis rate drops sharply, yielding $33{,}137$ children
with only $1.28\%$ diagnosed. Even with this artificial processing step the
censoring times remain realistic, as they derive from the original patient
trajectories.

For evaluation, we include only children in the test set with sufficiently
long follow-up, defined as trajectories extending beyond age 9. Children in
this filtered test set who have not received an autism diagnosis are treated
as ground-truth non-susceptibles. Because this cohort has a very low event
rate, we report average precision (AP) alongside AUC-ROC in Figure
\ref{fig:asd}; AP is not reported for the simulation and diabetes
experiments, where the classes are approximately balanced by construction.

\begin{figure*}[!t]
    \centering
    \includegraphics[width=\textwidth]{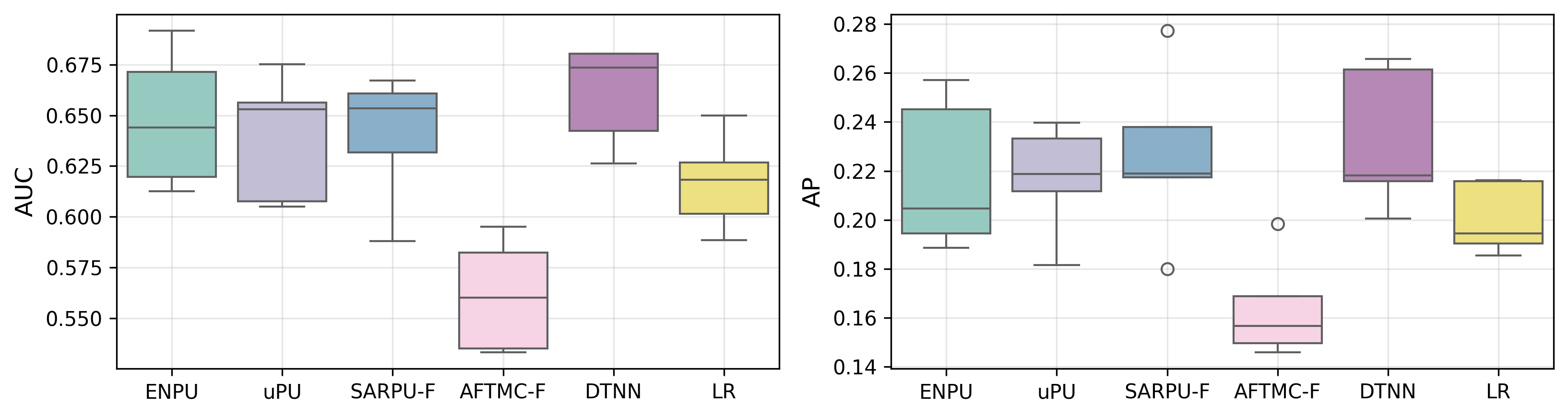}
    \caption{The boxplots show the AUC and AP scores for different methods used in the autism screening study. All methods are evaluated across 5 random seeds.}
    \label{fig:asd_full}
\end{figure*}

% Last section of the appendix: a heading plus one full-width figure. A
% double-column float (figure*) can only be placed at the top of a page that
% begins after the float is read, so it drifts away from its heading.
% \onecolumn starts a fresh page and flushes pending floats, so [!ht] then
% places the figure directly under its heading. Nothing follows this section,
% so the column change is not visible.
\onecolumn
\section{Methods Recommendation}\label{app: flow}
\begin{figure}[!ht]
    \centering
    \includegraphics[width=0.7\linewidth]{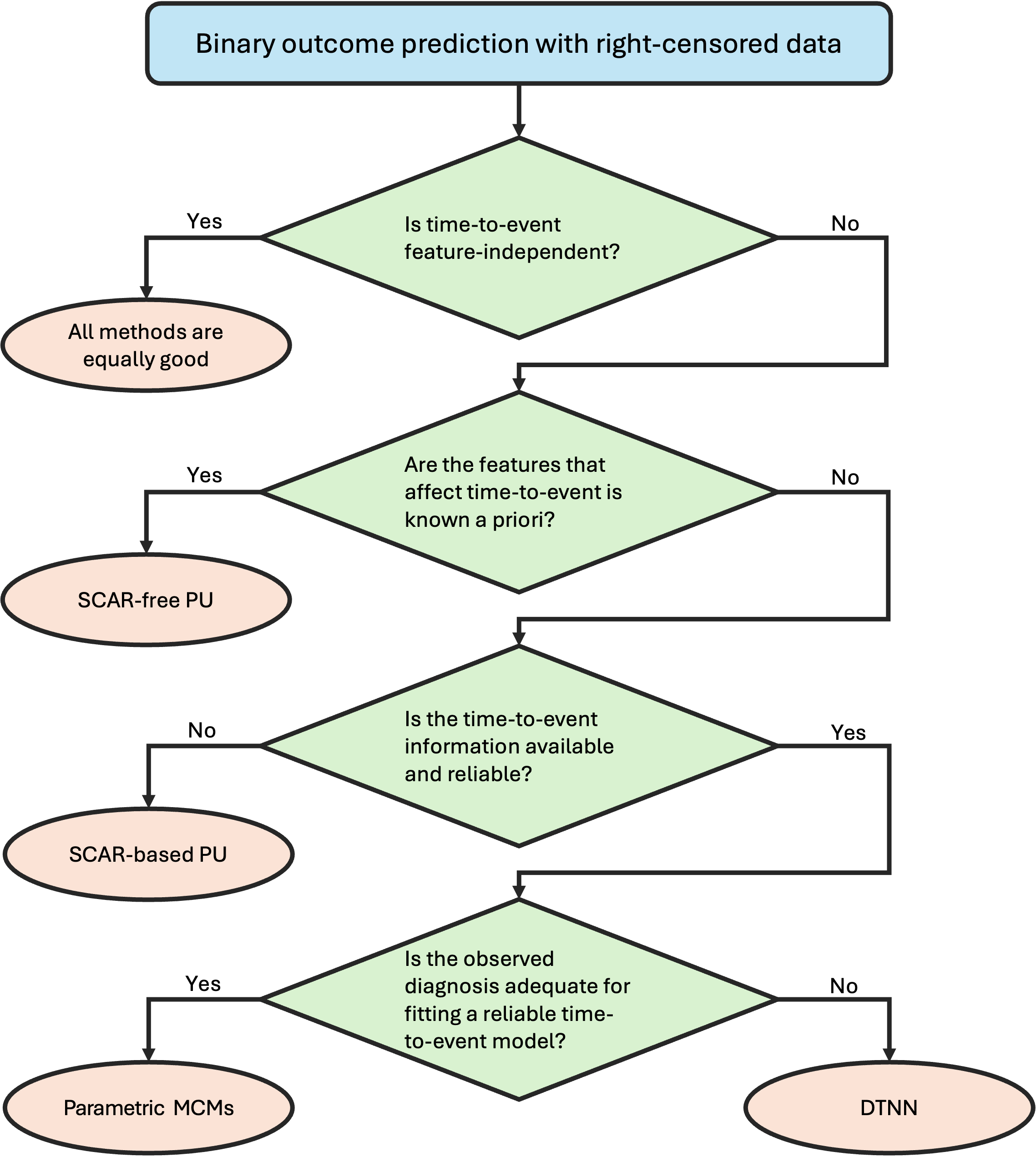}
    \caption{Methods recommendation flowchart for binary clinical outcome prediction with right-censored data.}
    \label{fig: Reccomend}
\end{figure}

\end{document}